\documentclass{article} 
\usepackage{nips14submit_e,times}
\usepackage{hyperref}
\usepackage{url}

\usepackage{subfigure}
\usepackage{graphicx}

\usepackage{amsmath,amsfonts,amsthm}
\usepackage[numbers,sort]{natbib}

\usepackage{algorithm}
\usepackage{algorithmic}
\usepackage{wrapfig}

\newcommand{\defeq}{:=}

\newcommand{\R}{\mathbb{R}}

\newcommand{\bzero}{\mathbf{0}}

\newcommand{\br}{\mathbf{r}}
\newcommand{\brt}{\mathbf{r}_{\boldsymbol\theta}}
\newcommand{\brp}{\mathbf{r}_{\boldsymbol\phi}}

\newcommand{\bw}{\mathbf{w}}

\newcommand{\bx}{\mathbf{x}}
\newcommand{\by}{\mathbf{y}}
\newcommand{\bz}{\mathbf{z}}

\newcommand{\bI}{\mathbf{I}}

\newcommand{\bY}{\mathbf{Y}}
\newcommand{\bX}{\mathbf{X}}
\newcommand{\cI}{\mathcal{I}}
\newcommand{\cD}{\mathcal{D}}
\newcommand{\cN}{\mathcal{N}}
\newcommand{\cO}{\mathcal{O}}
\newcommand{\cX}{\mathcal{X}}

\newcommand{\btheta}{\boldsymbol\theta}

\newcommand{\bphi}{\boldsymbol\phi}
\newcommand{\bLambda}{\boldsymbol\Lambda}

\newcommand{\Exp}{\mathbb{E}}

\newcommand{\sfmhmcf}{SSHMC}
\newcommand{\sfmhmc}{SSHMC}

\makeatletter
\newcommand{\shorteq}{%
 \settowidth{\@tempdima}{-}
 \,\resizebox{\@tempdima}{\height}{=} \,%
}
\title{Semi-Separable Hamiltonian Monte Carlo \\ 
           for Inference in Bayesian Hierarchical Models}

\author{
Yichuan Zhang \\
School of Informatics\\
University of Edinburgh\\
\texttt{Y.Zhang-60@sms.ed.ac.uk} \\
\And
Charles Sutton\\
School of Informatics\\
University of Edinburgh\\
\texttt{c.sutton@inf.ed.ac.uk} \\
}

\nipsfinalcopy 

\begin{document}

\maketitle

\begin{abstract}
Sampling from hierarchical Bayesian models is often difficult
for MCMC methods, because of the strong correlations between
the model parameters and the hyperparameters.
Recent Riemannian manifold Hamiltonian Monte Carlo (RMHMC) methods
have significant potential advantages in this setting, 
but are computationally expensive.  We introduce a new RMHMC method,
which we call \emph{semi-separable Hamiltonian Monte Carlo},
which uses a specially designed mass matrix that allows
the joint Hamiltonian over model parameters and
hyperparameters to decompose into two simpler Hamiltonians.
This structure is exploited by a new integrator which we call the
\emph{alternating blockwise leapfrog algorithm}.
The resulting method can mix faster
than simpler Gibbs sampling while being simpler and more
efficient than previous instances of RMHMC.
\end{abstract}

\section{Introduction}

Bayesian statistics provides  a natural way to manage model complexity and control
overfitting, with modern problems involving complicated models with a large number 
of parameters.
One of the most powerful advantages of the Bayesian approach
is hierarchical modeling, 
which allows partial pooling across a group of datasets,
allowing groups with little data to borrow information
from similar groups with larger amounts of data.
However, such models pose problems for Markov chain Monte Carlo (MCMC) methods,
because the joint posterior distribution is often pathological due to strong correlations between the model
parameters and the hyperparameters \cite{mark13}.
For example, one of the most powerful MCMC methods is
Hamiltonian Monte Carlo (HMC).
However, for hierarchical models even the mixing speed of HMC can be unsatisfactory in practice,
as has been noted several times in the literature \cite{mark13,
choo2000learning, radford2010}.
Riemannian manifold Hamiltonian Monte Carlo (RMHMC) \cite{mark11} is a recent
extension of HMC that aims to efficiently sample from challenging
posterior distributions by exploiting local geometric properties of the distribution of
interest.
However, it is computationally too expensive to be applicable to large scale
problems.

In this work, we propose a simplified RMHMC method, called Semi-Separable
Hamiltonian Monte Carlo (SSHMC), in which
the joint Hamiltonian over parameters and hyperparameters has
special structure, which we call \emph{semi-separability}, 
that allows it to be decomposed into two simpler, separable Hamiltonians.
This condition allows for a new efficient algorithm
which we call the \emph{alternating blockwise leapfrog algorithm}.
Compared to Gibbs sampling, SSHMC can make significantly larger moves in
hyperparameter space due to shared terms between the two simple Hamiltonians.
Compared to previous RMHMC methods, SSHMC yields simpler
and more computationally efficient samplers for many practical Bayesian
models.

\section{Hierarchical Bayesian Models}

Let $\cD = \{\cD_i\}_{i=1}^N$ be a collection of data groups
where $i$th data group is a collection of iid observations $\by_j =
\{y_{ji}\}_{i=1}^{N_i}$ and their inputs $\bx_j =
\{\bx_{ji}\}_{i=1}^{N_i}$. 
We assume the data follows a parametric distribution $p(\by_i \vert \bx_i,
\btheta_i)$, where $\btheta_i$ is the model parameter for group $i$.
The parameters are assumed
to be drawn from a prior $p(\btheta_i \vert \bphi)$, where $\bphi$ is the
hyperparameter with prior distribution $p(\bphi)$.
The joint posterior over model parameters
$\btheta = (\btheta_1, \ldots, \btheta_N)$ and hyperparameters $\bphi$ is then
\begin{eqnarray}
	p(\btheta, \bphi \vert \cD) \propto \prod_{i=1}^N p(\by_i \vert \bx_i,
	\btheta_i)p(\btheta_i \vert \bphi) p(\bphi).\label{eq:hbayes}
\end{eqnarray}
This \emph{hierarchical Bayesian} model is popular because
the parameters $\btheta_i$ for each group are coupled, allowing the groups to share statistical strength.
However, this property causes difficulties when approximating
the posterior distribution. In the posterior, the model parameters and hyperparameters are strongly correlated.
In particular, $\bphi$ usually controls the variance of $p(\btheta \vert \bphi)$ 
to promote partial pooling,
so the variance of $\btheta \vert \bphi, \cD$ depends strongly on $\bphi.$
This causes difficulties for many
MCMC methods, such as the Gibbs sampler and HMC.
An illustrative example of pathological structure in hierarchical models is 
the Gaussian funnel distribution \cite{radford2010}.
Its density function is defined as
$p(\bx, v) = \prod_{i=1}^n\cN(x_i \vert 0, e^{-v})\cN(v \vert 0, 3^2)$,
where $\bx$ is the vector of low-level parameters and $v$ is the variance
hyperparameters.
The pathological correlation between $\bx$ and $v$ is illustrated by
Figure \ref{funnel_ET}.

\section{Hamiltonian Monte Carlo on Posterior Manifold}
Hamiltonian Monte Carlo (HMC) is a gradient-based MCMC method with auxiliary
variables.
To generate samples from a target density $\pi(\bz)$, 
HMC constructs an ergodic Markov
chain with the invariant distribution $\pi(\bz, \br) = \pi(\bz)
\pi(\br)$, where $\br$ is an auxiliary variable.
The most common choice of $\pi(\br)$ is a Gaussian distribution $ \cN(\bzero,
G^{-1})$ with precision matrix $G$.
Given the current sample $\bz$, the transition kernel of the HMC chain includes
three steps: first sample $\br \sim \pi(\br)$, second
propose a new sample $(\bz', \br')$ by simulating the Hamiltonian dynamics
and finally accept the proposed sample with probability $\alpha = \min\left\{1,
\pi(\bz',\br') / \pi(\bz, \br)\right\}$, otherwise leave $\bz$
unchanged.
The last step is a Metropolis-Hastings (MH) correction.
Define $H(\bz,\br) \defeq -\log \pi(\bz, \br)$.
The Hamiltonian dynamics is defined by the differential equations
$\dot{\bz} = \partial_{\br}{H}\quad
\dot{\br} = -\partial_{\bz}{H}$,
where $\bz$ is called the \emph{position} and $\br$ is called the \emph{momentum}.

It is easy to see that $\dot{H}(\bz, \br) = \partial_{\bz}{H}\dot{\bz} + 
\partial_{\br}{H}\dot{\br} = 0$, which is called the energy preservation property
\cite{radford2010, leimkuhler2004simulating}.
In physics, $H(\bz,\br)$ is known as the \emph{Hamiltonian energy},
and is decomposed into the sum of the \emph{potential energy} $U(\bz) \defeq -\log \pi(\bz)$ 
and the \emph{kinetic energy} $K(\br) \defeq -\log \pi(\br)$.
The most used discretized simulation in HMC is the \emph{leapfrog} algorithm,
which is given by the recursion
\begin{subequations} 
\begin{align}
	\br(\tau + \epsilon/2)&= \br(\tau) -
	\frac{\epsilon}{2} \nabla_{\bz}U(\tau) \label{LF1} \\
	\bz(\tau + \epsilon)&= \bz(\tau) + 
	\epsilon \nabla_{\br}K(\tau + \epsilon/2)
	\label{LF2} \\
	\br(\tau + \epsilon)&= \br(\tau + \epsilon/2) -
	\frac{\epsilon}{2}\nabla_{\btheta}U(\tau+\epsilon),
	\label{LF3}
	\vspace{-10pt}
\end{align}
\end{subequations} 
where $\epsilon$ is the step size of discretized simulation time. 
After $L$ steps from the current sample $(\bz(0),
\br(0)) = (\bz, \br)$, the new sample is proposed as the last point
$(\bz', \br')=(\bz(L\epsilon), \br(L\epsilon))$.
In Hamiltonian dynamics, the matrix $G$ is called the \emph{mass matrix}.
If $G$ is constant w.r.t. $\bz$, then $\bz$ and $\br$ are independent in
$\pi(\bz,\br)$.
In this case we say that $H(\bz,\br)$ is a \emph{separable} Hamiltonian.
In particular, we use the term \emph{standard HMC}
to refer to HMC using the identity matrix as $G$.
Although HMC methods often outperform other popular MCMC methods,
they may mix slowly if there are strong correlations between variables in
the target distribution.
Neal \cite{radford2010} showed that
HMC can mix faster if $G$ is not the identity matrix. Intuitively,
such a $G$ acts like a preconditioner.
However, if the curvature of $\pi(\bz)$ varies greatly, a global
preconditioner can be inadequate.

For this reason, recent work, notably that on Riemannian manifold HMC (RMHMC) 
\cite{mark11}, has considered
\emph{non-separable} Hamiltonian methods, in which 
$G(\bz)$ varies with position $\bz$,
so that  $\bz$ and $\br$ are no longer independent in $\pi(\bz,\br)$.  The resulting Hamiltonian $H(\bz,\br) = -\log \pi(\bz,\br)$ 
is called a \emph{non-separable} Hamiltonian.
For example,  for Bayesian inference problems, Girolami et al. 
\cite{mark11} proposed using the Fisher Information Matrix
(FIM) of $\pi(\btheta)$, which is the metric tensor of posterior manifold.
However, for a non-separable Hamiltonian, the simple leapfrog dynamics \eqref{LF1}-\eqref{LF3} do not yield a valid MCMC method, as they are
no longer reversible.
Simulation of general non-separable systems requires
the generalized leapfrog integrator (GLI) \cite{mark11}, which requires computing
higher order derivatives to solve a system of non-linear differential equations.
The computational cost of GLI in general is $\cO(d^3)$
where $d$ is the number of parameters, which is prohibitive for large $d$.

In hierarchical models, 
there are two ways to sample the posterior using HMC.
One way is to sample the joint posterior $\pi(\btheta, \bphi)$ directly.
The other way is to sample the conditional $\pi(\btheta \vert \bphi)$ and 
$\pi(\bphi \vert \btheta)$, simulating from each conditional distribution
using HMC. This strategy is called HMC within Gibbs \cite{radford2010}.
In either case, HMC chains tend to mix slowly in hyperparameter space,
because the huge variation of potential energy
across different hyperparameter values can easily overwhelm the kinetic energy in separable 
HMC \cite{radford2010}.
Hierarchical models also pose a challenge to RMHMC,
if we want to sample the model parameters and hyperparameters jointly.
In particular, the closed-form FIM of the joint posterior $\pi(\btheta,
\bphi)$ is usually unavailable.
Due to this problem, even sampling some toy models like the Gaussian
funnel using RMHMC becomes challenging.
Betancourt \cite{MJ12} proposed a new metric that uses a transformed 
Hessian matrix of $\pi(\btheta)$,
and Betancourt and Girolami \cite{mark13}
demonstrate the power of this method for efficiently sampling
hyperparameters of hierarchical models on some simple benchmarks like Gaussian
funnel.
However, the transformation 
requires computing eigendecomposition of
the Hessian matrix, which is infeasible in high dimensions.

Because of these technical difficulties, 
RMHMC for hierarchical models is usually used within a block Gibbs sampling scheme, 
alternating between $\btheta$ and $\bphi$.  
This \emph{RMHMC within Gibbs} strategy is useful because the simulation of the non-separable
dynamics for the conditional distributions may have much lower computational cost than
that for the joint one.
However, as we have discussed, in hierarchical models these variables
tend be very strongly correlated, and
it is well-known that Gibbs samplers mix slowly in such cases \cite{robert2004monte}.
So, the Gibbs scheme limits the true power of RMHMC.

\section{Semi-Separable Hamiltonian Monte Carlo}

In this section we propose a \emph{non-separable} HMC method that does not have
the limitations of Gibbs
sampling and that scales to relatively high dimensions,
based on a novel property that we will call semi-separability.
We introduce new
HMC methods that rely on semi-separable Hamiltonians,
which we call \emph{semi-separable Hamiltonian Monte Carlo (SSHMC)}.

\subsection{Semi-Separable Hamiltonian}

In this section, we define the semi-separable Hamiltonian system.
Our target distribution will be the posterior 
$\pi(\btheta, \bphi) = \log p(\btheta,\bphi | \cD)$ 
of a hierarchical model \eqref{eq:hbayes},
where $\btheta \in \R^n$ and $\bphi \in \R^m$.
Let $(\brt,\brp) \in \R^{m+n}$ be the vector of momentum variables corresponding
to $\btheta$ and $\bphi$ respectively.
The non-separable Hamiltonian is defined as
\begin{eqnarray}
	\label{ham0}
	H(\btheta, \bphi, \brt, \brp) = U(\btheta, \bphi) + K(\brt, 
	\brp \vert \btheta, \bphi),
\end{eqnarray}
where the potential energy is $U(\btheta, \bphi)=-\log \pi(\btheta, \bphi)$ 
and the kinetic energy is
$K(\brt, \brp \vert \btheta, \bphi) = -\log \cN(\brt,\brp; \bzero,
G(\btheta, \bphi)^{-1}),$ which includes the normalization term 
$\log \vert G(\btheta, \bphi) \vert$. 
The mass matrix $G(\btheta, \bphi)$ can be an arbitrary p.d. matrix.
For example, previous work on RMHMC \cite{mark11} has chosen
$G(\btheta, \bphi)$ to be FIM of the joint posterior
$\pi(\btheta, \bphi)$, resulting in an HMC method that requires
\emph{$\cO\left( \left(  m + n\right)^3 \right)$} time.
This limits applications of RMHMC to large scale problems.

To attack these computational challenges, we introduce
restrictions on the mass matrix $G(\btheta, \bphi)$
to enable efficient simulation.
In particular, we restrict $G(\btheta, \bphi)$ to have the form
\[G(\btheta, \bphi) = \left( \begin{array}{ccc}
G_{\btheta}(\bphi, \bx) & \bzero  \\
\bzero & G_{\bphi}(\btheta) \end{array} \right),\]
where $G_{\btheta}$ and $G_{\bphi}$ are the precision matrices of $\brt$ and
$\brp$, respectively. Importantly, we restrict $G_{\btheta}(\bphi, \bx)$ to be
independent of $\btheta$ and $G_{\bphi}(\btheta)$ to be independent of $\bphi$. 
If $G$ has these properties, we call the resulting Hamiltonian a 
\emph{semi-separable} Hamiltonian. A semi-separable Hamiltonian is still
in general non-separable, as the two random vectors 
$(\btheta, \bphi)$ and $(\brt,\brp)$ are not independent.

The semi-separability property has important computational advantages.
First, because $G$ is block diagonal, the cost of
matrix operations reduces from $\cO((n+m)^k)$ to $\cO(n^k)$.
Second, and more important,
substituting the restricted mass matrix into \eqref{ham0}
results in the potential and kinetic energy:
\begin{align}
	\label{pot}
	U(\btheta, \bphi) 
	&= -\sum_{i} [\log p(\by_i \vert \btheta_i, \bx_i)
	    + \log p(\btheta_i \vert \bphi)]
 	    -\log p(\bphi),\\
	K(\brt, \brp \vert \bphi, \btheta) 
	&= \frac{1}{2}[ \brt^T G_{\btheta}(\bx, \bphi)\brt 
	    + \brp^T G_{\bphi}(\btheta) \brp
            + \log \left\vert G_{\btheta}(\bx,
    		\bphi) \right\vert
	    + \log \left\vert G_{\bphi}(\btheta) \right\vert].
	\label{kin}
\end{align}

If we fix $(\btheta, \brt)$ or $(\bphi, \brp)$, the non-separable
Hamiltonian \eqref{ham0} can be seen as a separable Hamiltonian plus some 
constant terms. 
In particular, define the notation
$$A(\brt \vert \bphi) = \frac{1}{2} \brt^T G_{\btheta}(\bx, \bphi)\brt,
\qquad A(\brp \vert \btheta) = \frac{1}{2} \brp^T G_{\bphi}(\btheta)\brp.$$
Then, considering $(\bphi,\br_{\bphi})$ as fixed, the non-separable Hamiltonian $H$
in \eqref{ham0}
is different from the following separable Hamiltonian 
\begin{eqnarray}
	\label{ham1}
	H_1(\btheta, \brt) &=&  U_1(\btheta \vert \bphi, \brp) + K_1(\brt \vert
	\bphi),\\ 
	U_1(\btheta \vert \bphi, \brp) &=& 
	-\sum_{i} [\log p(\by_i \vert \btheta_i, \bx_i)
	    + \log p(\btheta_i \vert \bphi)]
	    + A(\brp \vert \btheta) 
	    +\frac{1}{2} \log \left\vert G_{\bphi}(\btheta) \right\vert,\label{h1b}\\
	K_1(\brt \vert \bphi)&=& A(\brt \vert \bphi) 
    \end{eqnarray}
only by some constant terms that do not depend on $(\btheta, \brt)$.
What this means is that any update to $(\btheta, \brt)$
that leaves $H_1$ invariant leaves the joint Hamiltonian
$H$ invariant as well.  An example is the leapfrog dynamics on $H_1$,
where $U_1$ is considered the potential energy, and $K_1$ the kinetic energy.


Similarly, if $(\btheta, \brt)$ are fixed, then $H$
differs from the following separable Hamiltonian 
\begin{eqnarray}
	\label{ham2}
	H_2(\bphi, \br_{\bphi}) &=&  U_2(\bphi \vert \btheta, \brt) + K_2(\brp \vert
	\btheta),\\
	U_2(\bphi \vert \btheta, \brt) &=& 
	-\sum_i \log p(\btheta_i \vert\bphi)
	- \log p(\bphi) 
	+ A(\brt \vert \bphi) 
	+ \frac{1}{2} \log \left\vert
	G_{\btheta}(\bx, \bphi)\right\vert,\label{h2b}\\
	K_2(\brp \vert \btheta) &=& A(\brp \vert \btheta) 
\end{eqnarray}
only by terms that are constant with respect to $(\bphi, \brp).$

Notice that $H_1$ and $H_2$ are coupled by the terms 
$A(\brt \vert \bphi)$ and $A(\brp \vert \btheta)$. Each of these terms
appears in the kinetic
energy of one of the separable Hamiltonians, but in the potential energy of the other one.
We call these terms \emph{auxiliary potentials} because they are potential energy terms introduced by the auxiliary variables. 
These auxiliary potentials
are key to our method (see Section~\ref{sec:gibbs}).

\begin{wrapfigure}{R}{0.5\textwidth}
		\centering
		\includegraphics[width=0.5\textwidth]{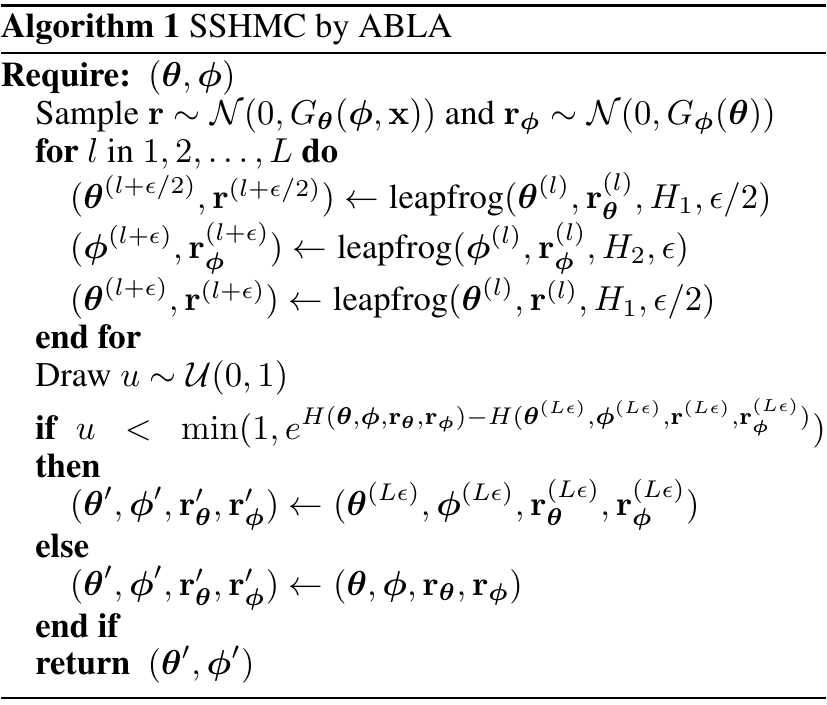}
		\vspace{-20pt}
\end{wrapfigure}

\subsection{Alternating block-wise leapfrog algorithm}
Now we introduce an efficient SSHMC method that exploits the semi-separability property.
As described in the previous section, 
any update to $(\btheta, \brt)$
that leaves $H_1$ invariant also leaves the joint Hamiltonian
$H$ invariant, as does any update to $(\bphi, \brp)$ that
leaves $H_2$ invariant. So a natural idea is simply to alternate
between simulating the Hamiltonian dynamics for $H_1$ 
and that for $H_2$. Crucially, even though the total Hamiltonian
$H$ is not separable in general,
both $H_1$ and $H_2$ \emph{are} separable.
Therefore when simulating $H_1$ and $H_2$, the 
simple leapfrog method can be used, and the more complex GLI method
is not required.

We call this method the \emph{alternating block-wise leapfrog algorithm} (ABLA),
shown in Algorithm 1.  In this figure the function ``leapfrog'' returns
the result of the leapfrog dynamics \eqref{LF1}-\eqref{LF3} for the given
starting point, Hamiltonian, and step size.  We call each iteration
of the loop from $1 \ldots L$ an \emph{ALBA step}.
For simplicity, we have shown one leapfrog step for $H_1$ and $H_2$
for each ALBA step, but in practice it is useful to use multiple leapfrog
steps per ALBA step.
ABLA has discretization error due to the leapfrog discretization, 
so the MH correction is required.  If it is possible to
simulate $H_1$ and $H_2$ exactly, then $H$ is preserved
exactly and there is no need for MH correction.

To show that the SSHMC method by ALBA preserves the distribution $\pi(\btheta, \bphi)$, we also 
need to show the ABLA is a time-reversible and volume-preserving transformation
in the joint space of $(\btheta, \brt, \bphi, \brp)$.
Let $\cX = \cX_{\btheta, \brt} \times \cX_{\bphi, \brp}$ where 
$(\btheta, \brt) \in \cX_{\btheta, \brt}$ and 
$(\bphi, \brp) \in \cX_{\bphi, \brp}$.
Obviously, any reversible and volume-preserving transformation in a subspace of
$\cX$ is also reversible and volume-preserving in $\cX$.
It is easy to see that each leapfrog step in the ABLA algorithm is
reversible and volume-preserving in either $\cX_{\btheta, \brt}$ or
$\cX_{\bphi, \brp}$.
One more property of integrator of interest is \emph{symplecticity}. 
Because each leapfrog integrator is symplectic in the subspace of $\cX$
\cite{leimkuhler2004simulating}, they are also
symplectic in $\cX$.  
Then because ABLA is a composition of symplectic leapfrog integrators,
and the composition of symplectic transformations is symplectic,
we know ABLA is symplectic.

We emphasize that ABLA is 
actually \emph{not} a discretized simulation of the semi-separable
Hamiltonian system $H$, that is, if starting at a point
$(\btheta, \brt, \bphi, \brp)$ in the joint space, we run
the exact Hamiltonian dynamics for $H$ for a length of time $L$,
the resulting point will not be the same as that returned by ALBA
at time $L$ even if the discretized time step is infinitely small.
For example, ALBA simulates $H_1$ with step size $\epsilon_1$ and 
$H_2$ with step size $\epsilon_2$ where $\epsilon_1 = 2\epsilon_2$,
when $\epsilon_2\rightarrow 0$ that preserves $H$.

\subsection{Connection to other methods}\label{sec:gibbs}

Although the SSHMC method may seem similar to RMHMC within Gibbs (RMHMCWG),
SSHMC is actually very different.  The difference is in the last two terms of 
\eqref{h1b} and \eqref{h2b}; if these are omitted from SSHMC and
the Hamiltonians for $\pi(\btheta \vert \bphi)$, 
then we obtain HMC within Gibbs.
Particularly important among these two terms is the 
auxiliary potential, because
it allows each of the separable Hamiltonian
systems to \emph{borrow energy} from the other one.
For example, if the previous leapfrog step increases the kinetic energy
$K_1(\brt | \bphi)$ in $H_1(\btheta, \brt)$, then, in the next leapfrog step for $H_2(\bphi, \brp)$, 
we see that $\bphi$ will have greater \emph{potential} energy 
$U_2(\bphi | \btheta, \brt)$,
because the auxiliary potential $A(\brt | \bphi)$ is shared.
That allows the leapfrog step to accommodate a larger change of $\log p(\bphi
\vert \btheta)$ using $A(\brt | \bphi)$.
So, the chain will mix faster in $\cX_{\bphi}$.
By the symmetry of $\btheta$ and $\bphi$, the auxiliary potential will also
accelerate the mixing in $\cX_{\btheta}$.

Another way to see this is that 
the dynamics in RMHMCWG for $(\brp, \bphi)$ preserves the distribution 
$\pi(\btheta, \brp, \bphi) = \pi(\btheta, \bphi)\cN(\brp; \bzero 
, G_{\bphi}(\bphi)^{-1})$ 
but not the joint $\pi(\btheta, \bphi, \brt, \brp)$.
That is because the Gibbs sampler does not take into account
the effect of $\bphi$ on $\brt$.
In other words, the Gibbs step has the stationary
distribution $\pi(\bphi, \brp \vert \btheta)$ rather than
$\pi(\bphi, \brp \vert \btheta, \brt).$
The difference between 
the two is the auxiliary potential.
In contrast, the SSHMC methods preserve the Hamiltonian of  
$\pi(\btheta, \bphi, \brt, \brp)$.


\subsection{Choice of mass matrix}
The choice of $G_{\btheta}$ and $G_{\bphi}$ in SSHMC is usually similar to
RMHMCWG.
If the Hessian matrix of $-\log p(\btheta \vert \by, \bx, \bphi)$ is independent
of $\btheta$ and always p.d., it is natural to define $G_{\btheta}$ as the inverse 
of the Hessian matrix.
However, for some popular models, e.g., logistic regression, the Hessian
matrix of the likelihood function depends on the parameters $\btheta$.
In this case, one can use any approximate Hessian $B$, like the Hessian at the
mode, and define
$G_{\btheta} \defeq (B + B(\bphi))^{-1}$, where $B(\bphi)$ is the Hessian of the prior distribution.
Such a rough approximation is usually good enough to improve the mixing speed,
because the main difficulty is
the correlation between model parameters and hyperparameters.

In general, because the computational bottleneck in HMC and SSHMC 
is computing the gradient of the target distribution, both methods have the
same computational complexity $\cO(lg)$, where $g$ is the cost of computing
the gradient and $l$ is the total number of leapfrog steps per iteration.
However, in practice we find it very beneficial to use 
multiple steps in each blockwise leapfrog update in ALBA; this can cause
SSHMC to require more time than HMC.
Also, depending on the mass matrix $G_{\btheta}$, the
cost of leapfrog a step in ABLA may be different from those in standard HMC. 
For some choices of $G_{\btheta}$, the leapfrog step in ABLA
can be even faster than one leapfrog step of HMC.
For example, in many models the computational bottleneck is the gradient
$\nabla_{\bphi} \log Z(\bphi)$, $Z(\bphi)$ is the normalization in prior.
Recall that $G_{\theta}$ is a function of $\bphi$.
If $ \vert G_{\theta} \vert = Z(\bphi)^{-1}$, $Z(\bphi)$ will be canceled out,
avoiding computation of $\nabla_{\bphi} \log Z(\bphi)$.
One example is using $G_{\bx} = e^v \bI$ in Gaussian funnel distribution
aforementioned in Section 2.
A potential problem of such $G_{\theta}$ is that the
curvature of the likelihood function $p(\cD \vert \btheta)$ is ignored.
But when the data in each group is sparse and the parameters $\btheta$ are
strongly correlated, this $G_{\theta}$ can give
nearly optimal mixing speed and make SSHMC much faster.

In general, any choice of $G_{\btheta}$ and $G_{\bphi}$ that would be valid for 
separable HMC with Gibbs is also valid for SSHMC.

\section{Experimental Results}
In this section, we compare the performance of \sfmhmc\ with the standard HMC
and RMHMC within Gibbs \cite{mark11} on four benchmark models.\footnote{Our use
of a Gibbs scheme for RMHMC follows standard practice \cite{mark11}.}
The step size of all methods are manually tuned so that the acceptance rate is
around $70$-$85\%$.
The number of leapfrog steps are tuned for each method using preliminary runs.
The implementation of RMHMC we used is from \cite{mark11}.
The running time is wall-clock time measured after burn-in. 
The performance is evaluated by the minimum Effective Sample Size (ESS) over
all dimensions (see \cite{geyer1992practical}).
When considering the different computational complexity of methods, our main
efficiency metric is time normalized ESS.

\begin{figure} [t!]
\begin{center}
\begin{tabular}{llll}
\resizebox{0.22\textwidth}{!}{\includegraphics{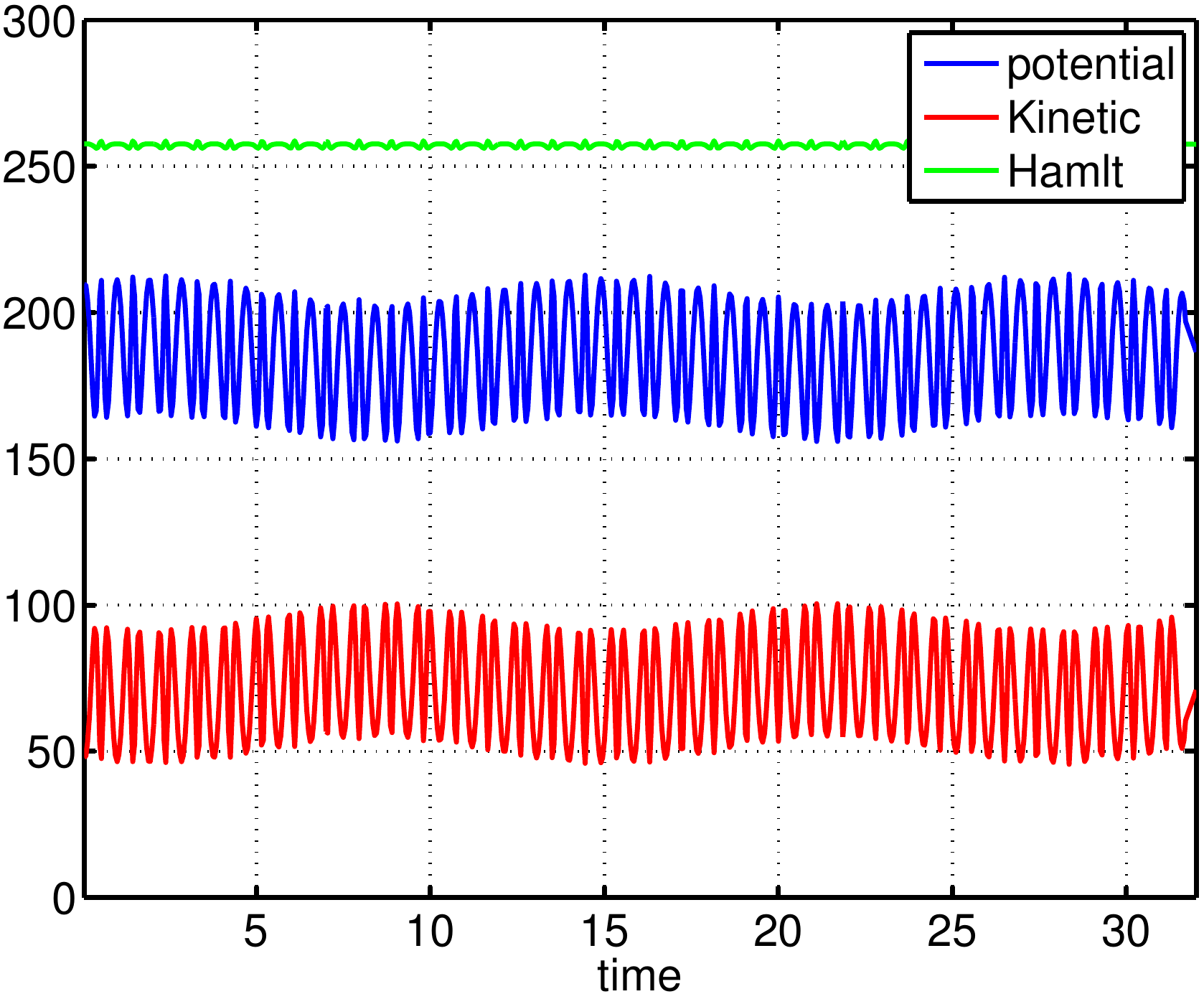}}
&
\resizebox{0.22\textwidth}{!}{\includegraphics{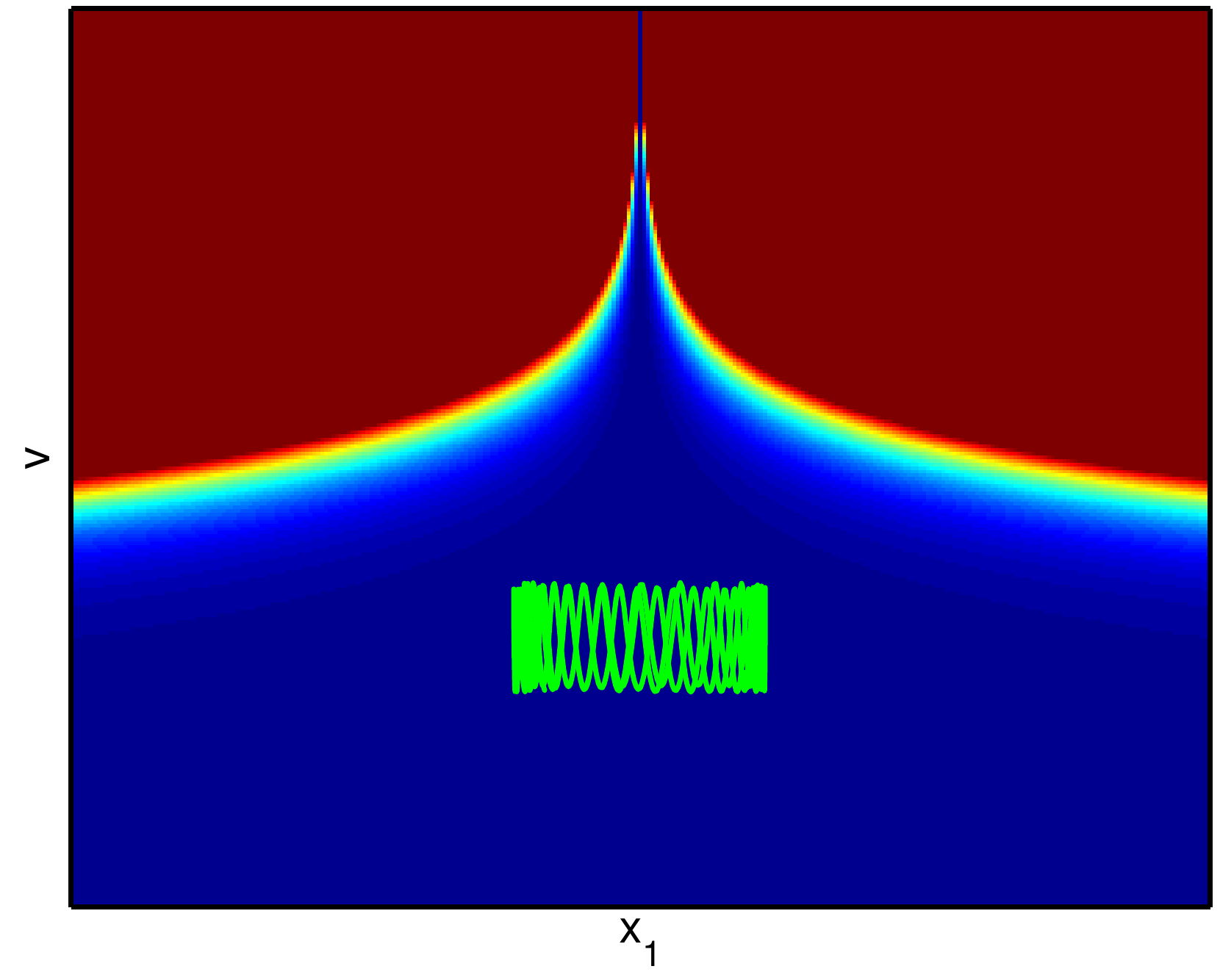}}
&
\resizebox{0.22\textwidth}{!}{\includegraphics{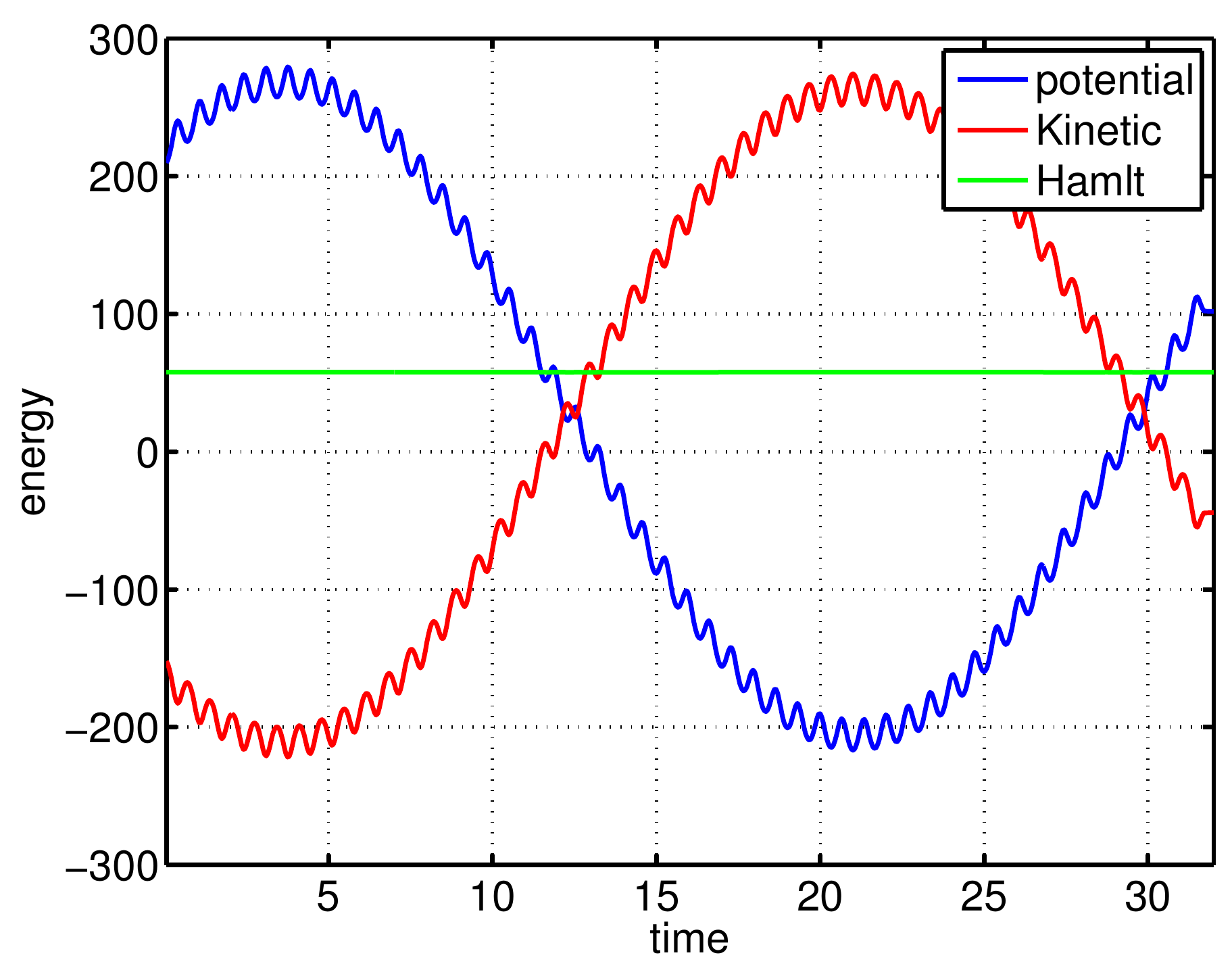}}
&
\resizebox{0.22\textwidth}{!}{\includegraphics{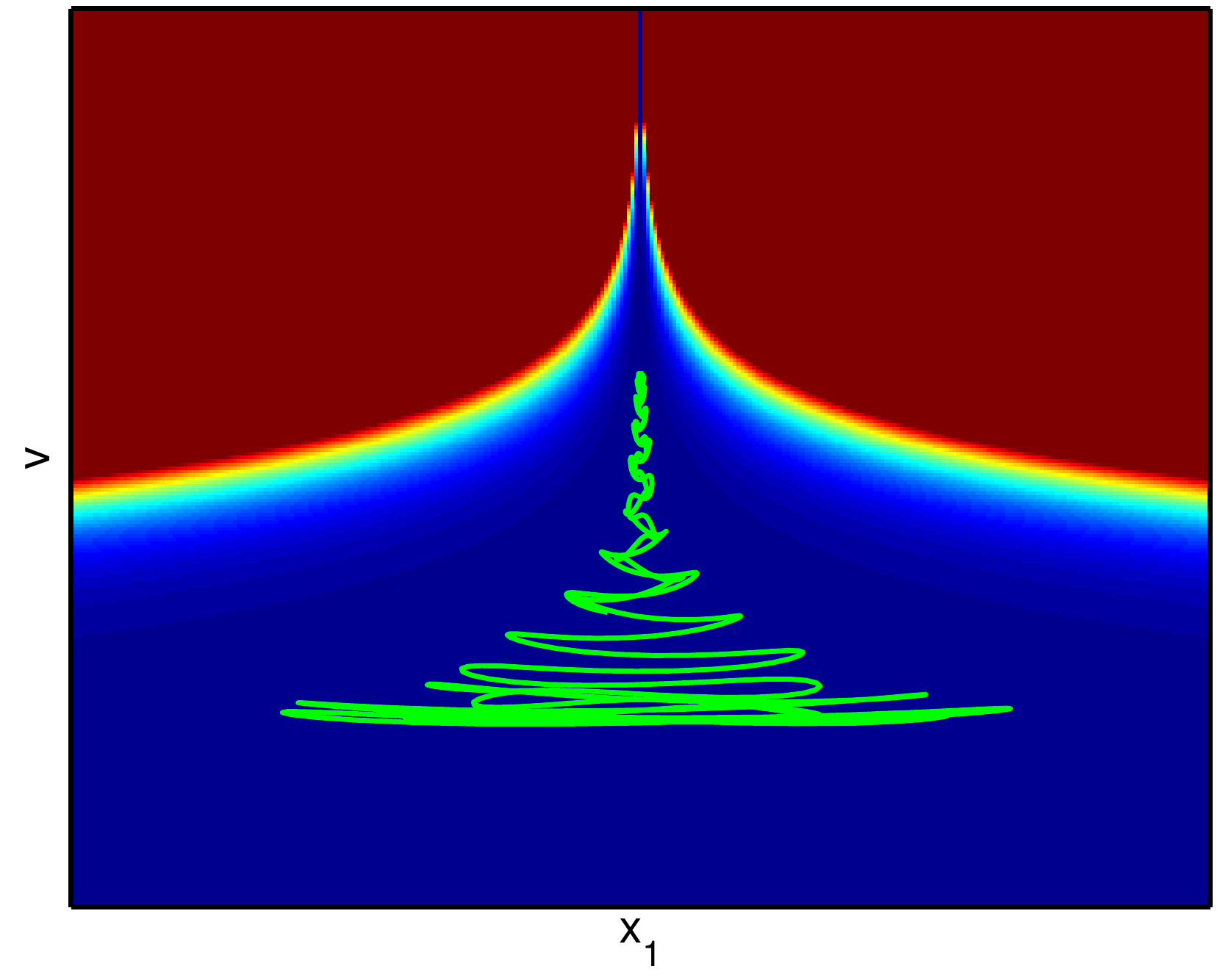}}\\
\multicolumn{2}{c}{HMC with diagonal constant mass} &
\multicolumn{2}{c}{SSHMC (semi-separable mass)}
\vspace{-10pt}
\end{tabular}
\caption{The trace of energy over the simulation time and
the trajectory of the first dimension of 100 dimensional Gaussian $\bx_1$
(vertical axis) and
hyperparameter $v$ (horizontal axis). 
The two simulations start with the same initial point
sampled from the Gaussian Funnel.
}\label{funnel_ET}
\end{center}
\vspace{-5pt}
\end{figure}

\begin{table}[tb!]
	\centering
	\begin{tabular}{lllll}
		 & time(s) & min ESS($\bx$, $v$) & min ESS/s ($\bx$, $v$)
		 &MSE($\Exp[v]$, $\Exp[v^2]$)\\
		 \hline
		 HMC & 5.16&(302.97, 26.30)&(58.64, 5.09)&(2.28, 1.34)\\
		RMHMC(Gibbs) &2.7& (2490.98, 8.93) &(\textbf{895.15},
		3.21)&(1.95, 1.33)\\
		\sfmhmcf\ &37.35& (\textbf{3868.79}, \textbf{1541.67})&(103.57,
		\textbf{41.27})&(\textbf{0.04}, \textbf{0.02})\\
		\hline
	\end{tabular}
	\caption{The result of ESS of 5000 samples on 100 + 1 dimensional
		Gaussian Funnel distribution.
		$\bx$ are model parameters and $v$ is the
		hyperparameter.
		The last column is the mean squared error of the
		sample estimated mean and variance of hyperparameter.}
	\label{tab:ESS_GF}
\end{table}

\begin{table}[tb!]
	\centering
	\begin{tabular}{lllll}
		 & running time(s) &ESS $\btheta$ (min, med, max)& ESS $v$ &
		 min ESS/s\\
		 \hline
		 HMC & 378&(2.05, 3.68, 4.79) $\times 10^3$ &815&2.15\\
		 RMHMC(Gibbs) &411&(0.8, \textbf{4.08}, \textbf{4.99})$\times
		 10^3$ &271&0.6\\
		\sfmhmcf\ &385.82&(\textbf{2.5}, 3.42, 4.27)$\times 10^3$
		&\textbf{2266}&\textbf{5.83}\\
		\hline
	\end{tabular}
	\caption{The results of ESS of 5000 samples after 1000 burn-in on Hierarchical
	Bayesian Logistic Regression. $\btheta$ are 200 dimensional model
	parameters and $v$ is the hyperparameter.}
		\label{tab:ESS_HBLR}
\end{table}

\begin{table}[tb!]
	\centering
	\begin{tabular}{lllll}
		 & time (s)& ESS $\bx$(min, med, max) & ESS$(\beta, \sigma,
		 \phi)$ & min ESS/s\\
		 \hline
		 HMC &162&(1.6, 2.2, 5.2)$\times 10^2$&(50, 50, 128)& 0.31\\
		 RMHMC(Gibbs) &183&(12.1, 18.4, 33.5)$\times 10^2$&(385, 163,
		411)&0.89\\
		 \sfmhmcf\ &883&(\textbf{78.4}, \textbf{98.9},
		 \textbf{120.7})$\times 10^2$ &(\textbf{4434}, \textbf{1706},
		 \textbf{1390})&\textbf{1.57}\\
		\hline
	\end{tabular}
	\caption{The ESS of 20000 posterior samples of Stochastic Volatility
		after 10000 burn-in. $\bx$ are latent volatilities over 2000 time
		lags and $(\beta, \sigma, \phi)$ are hyperparameters. Min ESS/s
		is the lowest ESS over all parameters normalized by running time.}
	\label{tab:ESS_StochV}
    \vspace{-15pt}
\end{table}

\begin{figure} [tb]
\begin{center}
\centering
\begin{tabular}{lll}

\resizebox{0.3\textwidth}{!}{\includegraphics{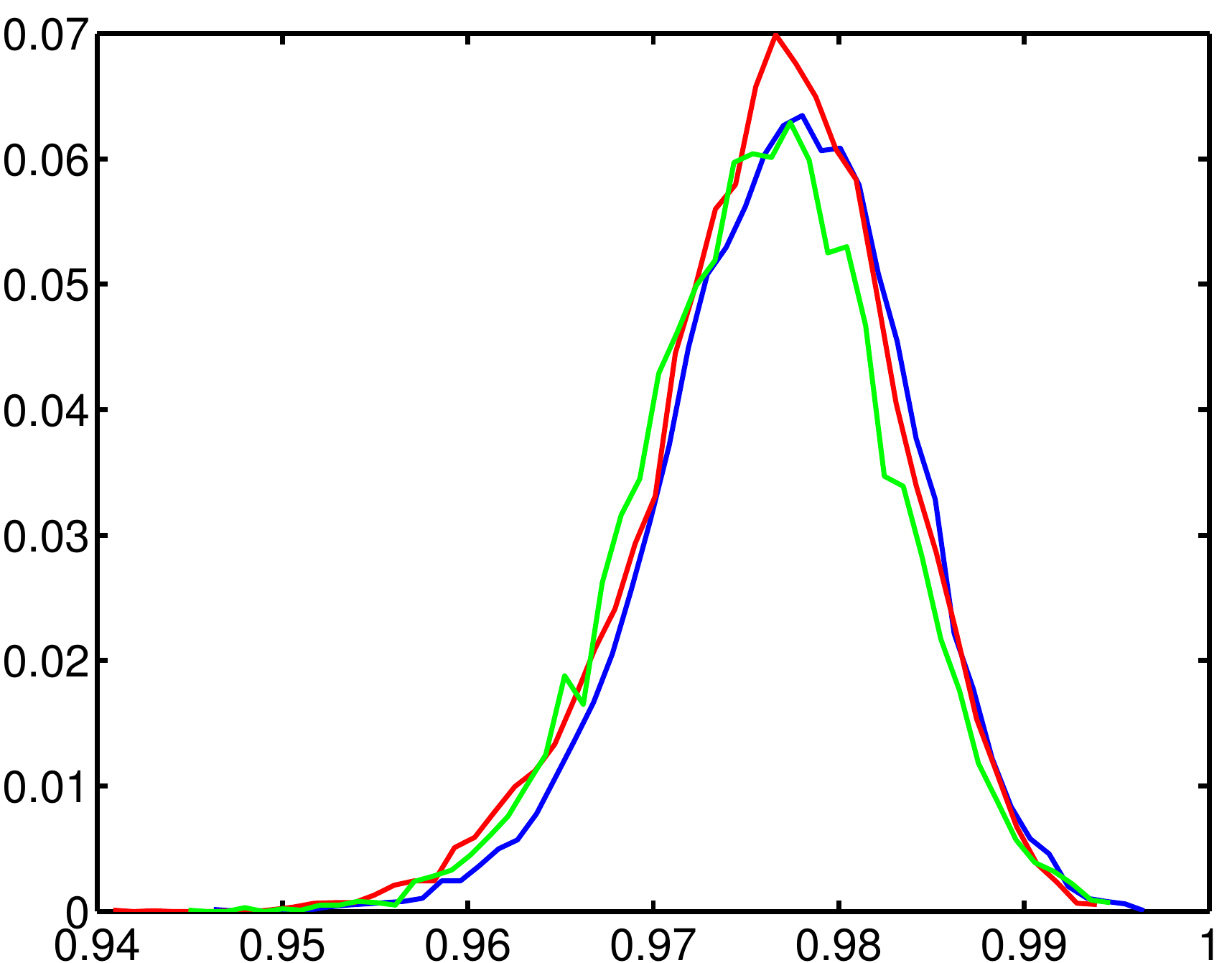}}
&
\resizebox{0.3\textwidth}{!}{\includegraphics{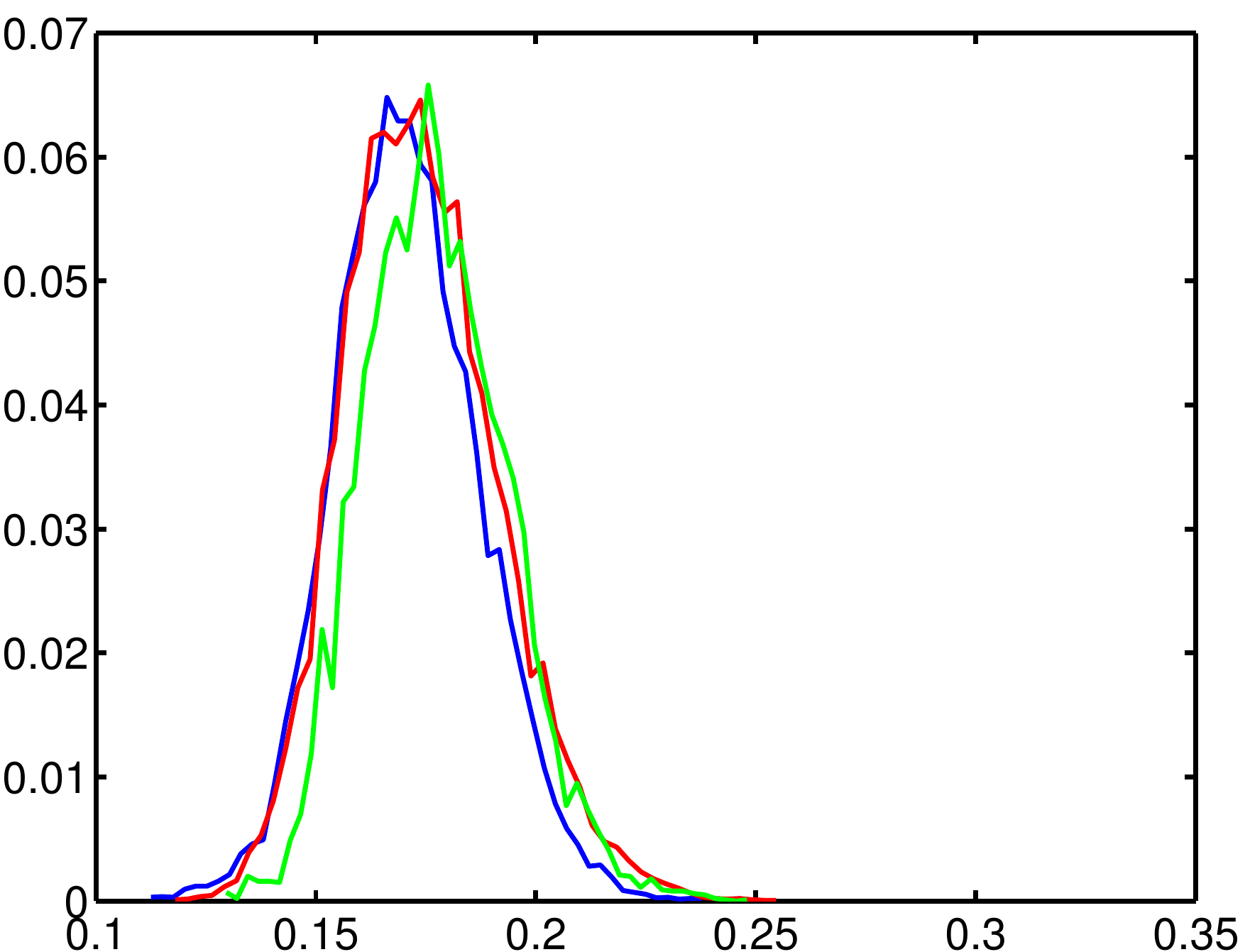}}
&
\resizebox{0.3\textwidth}{!}{\includegraphics{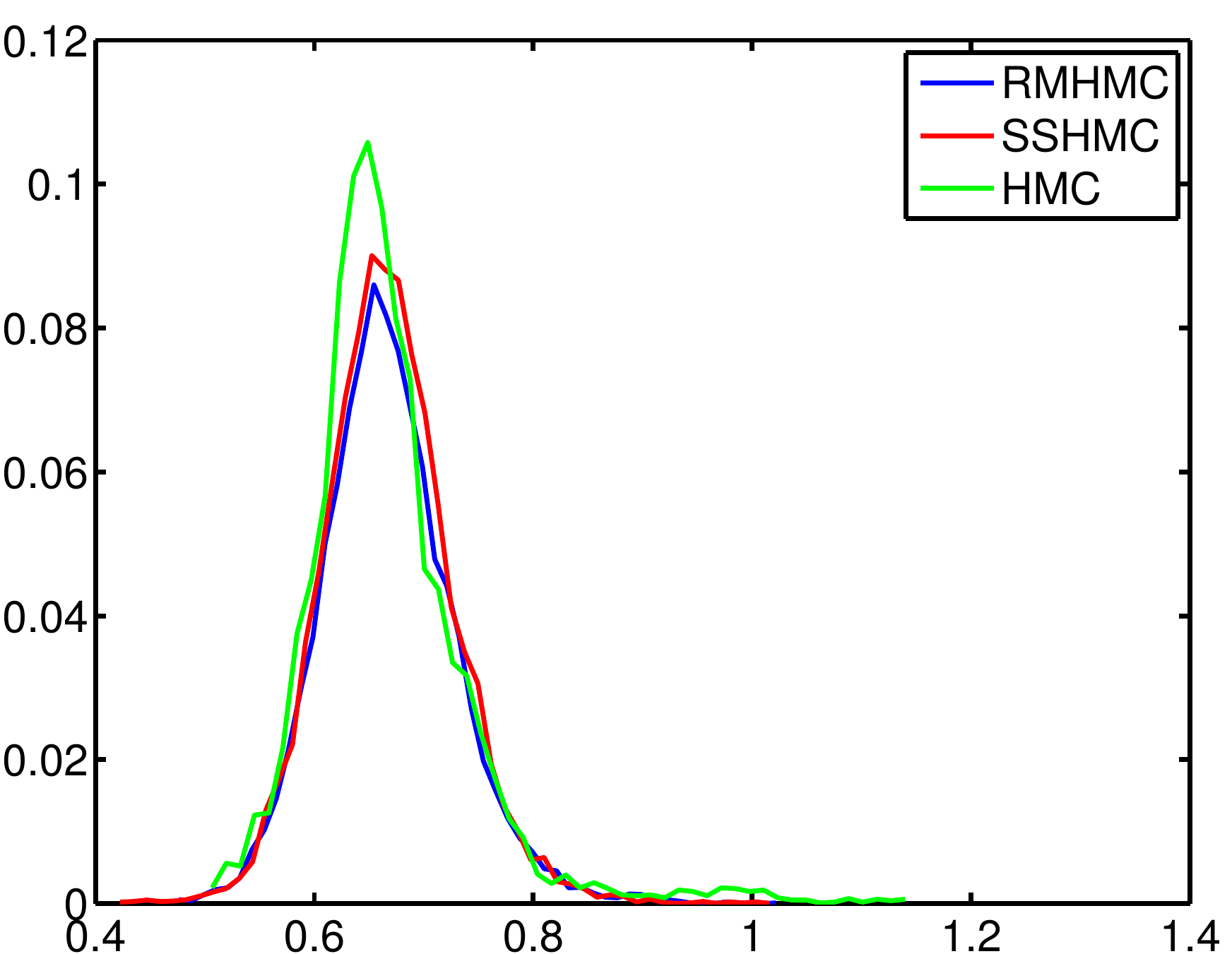}}

\end{tabular}
\caption{ The normalized histogram of 20000 posterior samples of
hyperparameters (from left to right $\phi$, $\sigma$, $\beta$) after 10000
burn-in samples. The data is generated by the hyperparameter $(\phi=0.98,
\sigma = 0.15, \beta = 0.65)$. It is clear that empirical distributions
by the three  methods are consistent. But, it is clear that SSHMC and RMHMC converges
faster than HMC.
\vspace{-15pt}}
\label{hist_hyper_SV}
\end{center}
\end{figure}

\subsection{Demonstration on Gaussian Funnel}
We demonstrate SSHMC by sampling the Gaussian Funnel (GF) defined in
Section 2.
We consider $n=100$ dimensional low-level parameters $\bx$ and 1
hyperparameter $v$. 
RMHMC within Gibbs on GF has block diagonal mass matrix defined as 
 $G_{\bx} = -\partial_v^2\log p(x, v)^{-1} = e^v \bI$ and
$G_v = -\Exp_{\bx}[\partial_v^2\log p(x, v)]^{-1} = (n + \frac{1}{9})^{-1}$. 
We use the same mass matrix in SSHMC, because it is semi-separable.
We use 2 leapfrog steps for low-level parameters and 1 leapfrog step for the
hyperparameter in ABLA and the same leapfrog step size for the two separable
Hamiltonians.

We generate 5000 samples from each method after 1000 burn-in iterations.
The ESS per second (ESS/s) and mean squared error (MSE) 
of the sample estimated mean and variance of the
hyperparameter are given in Table \ref{tab:ESS_GF}.
Notice that RMHMC is much more efficient for the low-level variable
because of the adaptive mass matrix with hyperparameter.
Figure~\ref{funnel_ET} illustrates a dramatic difference between HMC and SSHMC.
It is clear that HMC suffers from oscillation of the hyperparameter in a narrow
region. 
That is because the kinetic energy limits the change of hyperparameters
\cite{radford2010, mark13}.
In contrast, \sfmhmc\ has much wider energy variation and the trajectory spans a
larger range of hyperparameter $v$.
The energy variation of SSHMC is similar to the RMHMC with Soft-Abs metric
(RMHMC-Soft-Abs) reported in
\cite{MJ12}, an instance of general RMHMC without Gibbs.
But compared with \cite{MJ12}, each ABLA
step is about 100 times faster than each generalized leapfrog step and SSHMC can generate 
around \emph{2.5 times} more effective samples per second than RMHMC-Soft-Abs.
Although RHMC within Gibbs has better ESS/s on the low level
variables, its estimation of the mean and variance is biased, indicating that
the chain has not yet mixed. 
More important, Table \ref{tab:ESS_GF} shows that the samples generated by SSHMC 
give nearly unbiased estimates of the mean and variance of the hyperparameter,
 which neither of the other methods are able to do.

%

\subsection{Hierarchical Bayesian Logistic Regression}
In this experiment, we consider hierarchical Bayesian logistic regression
with exponential prior for the variance hyperparameter $v$, that is
\begin{eqnarray*}
	p(\bw, \phi \vert \cD) \propto
	\prod_{i}\prod_{j}\sigma(y_{ij}\bw_{i}^T\bx_{ij})\cN(\bw_i
	\vert \bzero, v\bI)\text{Exp}(v \vert \lambda),
\end{eqnarray*}
where $\sigma$ is the logistic function $\sigma(z) = 1 / (1 + \exp(-z))$ and
$(y_{ij}, \bx_{ij})$ is the $j$th data points the $i$th group.
We use the Statlog (German credit) dataset from \cite{Bache+Lichman:2013}.
This dataset includes 1000 data points and each data has $16$ categorical
features and $4$ numeric features.
Bayesian logistic regression on this dataset has been considered as a
benchmark for HMC \cite{mark11, Hoffman11}, but the previous work uses
only one group in their experiments.
To make the problem more interesting, we partition the dataset into $10$ groups
according to the feature \emph{Purpose}.
The size of group varies from 9 to 285.
There are $200$ model parameters ($20$ parameters for each group) and $1$ 
hyperparameter. 

We consider the reparameterization of the hyperparameter $\gamma = \log v$.
For RMHMC within Gibbs, the mass matrix for group $i$ is 
$G_i \defeq \cI(\bx, \btheta)^{-1},$ where
$\cI(\bx, \btheta)$ is the Fisher Information matrix for model parameter
$\bw_i$ and constant mass $G_v$.
In each iteration of the Gibbs sampler, 
each $\bw_i$ is sampled from by RMHMC using 6 generalized leapfrog steps and
$v$ is sampled using 6 leapfrog steps.
For SSHMC, $G_i \defeq \text{Cov}(\bx) + \exp(\gamma)\bI$ and the same
constant mass $G_v$.

The results are shown in Table \ref{tab:ESS_HBLR}. 
SSHMC again has much higher ESS/s than the other methods.

%
%

\subsection{Stochastic Volatility}
A stochastic volatility model we consider is studied in
\cite{kim1998stochastic}, in which the latent volatilities
are modeled by an auto-regressive AR(1) process such that the observations are $y_t
= \epsilon_t \beta \exp(x_t/2)$ with latent variable $x_{t+1} = \phi x_t + 
\eta_{t+1}$. We consider the distributions $x_1 \sim \cN(0,\sigma^2 /
(1-\phi^2))$, $\epsilon_t \sim \cN(0, 1)$ and $\eta_t\sim(0, \sigma^2)$. The
joint probability is defined as
\begin{eqnarray*}
	p(\by, \bx, \beta, \phi, \sigma) &=& \prod_{t=1}^Tp(y_t \vert x_t,
	\beta)p(x_1)\prod_{t=2}^Tp(x_t \vert x_{t-1}, \phi, \sigma)
	\pi(\beta) \pi(\phi)\phi(\sigma),
\end{eqnarray*}
where the prior $\pi(\beta)\propto 1/\beta$, $\sigma^2 \sim \text{Inv-}
\chi^2(10, 0.05)$ and $(\phi + 1)/2 \sim \text{beta}(20,
1.5)$. 
The FIM of $p(\bx \vert \alpha, \beta, \phi, \by)$ depends on the hyperparameters
but not $\bx$, but the FIM of $p(\alpha, \beta, \phi \vert \bx, \by)$ depends on
$(\alpha, \beta, \phi)$.
For RMHMC within Gibbs we consider FIM as the metric tensor following \cite{mark11}.
For SSHMC, we define $G_{\btheta}$ as inverse Hessian of 
$\log p(\bx \vert \alpha, \beta, \phi, \by)$, but $G_{\bphi}$ as an
identity matrix.
In each ABLA step, we use 5 leapfrog steps for updates of $\bx$ and 2 leapfrog steps for 
updates of the hyperparameters, so that 
the running time of SSHMC is about 7 times that of 
standard HMC.

We generate 20000 samples using each method after 10000 burn-in samples. 
The histogram of hyperparameters is shown in Figure \ref{hist_hyper_SV}. 
It is clear that all methods approximately converge to the same distribution.
But from Table \ref{tab:ESS_StochV}, we see that SSHMC generates almost 
\emph{two times} as many ESS/s as RMHMC within Gibbs.

\subsection{Log-Gaussian Cox Point Process}
The log-Gaussian Cox Point Process (LGCPP) is another popular testing benchmark 
\cite{christensen2005, mark11, Wang:2013}.
We follow the experimental setting of Girolami and Calderhead \cite{mark11}.
The observations $\bY = \{y_{ij}\}$ are counts at the location $(i,j)$, 
$i,j = 1,\dots, d$ on a regular spatial grid, which are conditionally 
independent given a latent intensity process $\bLambda = \{\lambda(i,j)\}$ 
with mean $m\lambda(i,j) = m\exp(x_{i,j})$, where $m = 1/ d^2$, 
$\bX = \{x_{i,j}\}$, $\bx = \text{Vec}(\bX)$ and $\by = \text{Vec}(\bY)$.
$\bX$ is assigned by a Gaussian process prior, with mean function $m(x_{i,j}) =
\mu\mathbf{1}$ and covariance function $\Sigma(x_{i,j}, x_{i',j'}) =
\sigma^2\exp(-\delta(i,i',j,j')/\beta d)$ where $\delta(\cdot)$ is the
Euclidean distance between $(i,j)$ and $(i',j')$. The log joint
probability is given by $\log p(\by, \bx \vert \mu, \sigma, \beta)=
	\sum_{i,j}y_{i,j}x_{i,j} - m\exp(x_{i,j}) - \frac{1}{2}(\bx-\mu\mathbf{1})^T
	\Sigma^{-1}(\bx-\mu\mathbf{1}).
$
We consider a $32 \times 32$ grid that has 1024 latent variables.
Each latent variable $x_{i,j}$ corresponds to a single observation $y_{i,j}$.


We consider RMHMC within Gibbs with FIM of the conditional posteriors. 
See \cite{mark11} for the definition of FIM. 
The generalized leapfrog steps are required
for updating $(\sigma, \beta)$, but only the leapfrog steps are required for
updating $\bx$.
Each Gibbs iteration takes 20 leapfrog steps for $\bx$ and 1 general leapfrog
step for $(\sigma, \beta)$. 
In SSHMC, we use $G_{\bx} = \Sigma^{-1}$ and $G_{(\sigma, \beta)} = \bI$.
In each ABLA step, the update of $\bx$ takes 2 leapfrog steps and the update of
$(\alpha, \beta)$ takes 1 leapfrog step.
Each SSHMC transition takes 10 ALBA steps.
We do not consider HMC on LGCPP, because it mixes extremely slowly for
hyperparameters.

\begin{figure} [tb]
\begin{center}
\centering
\begin{tabular}{llll}

\resizebox{0.22\textwidth}{!}{\includegraphics{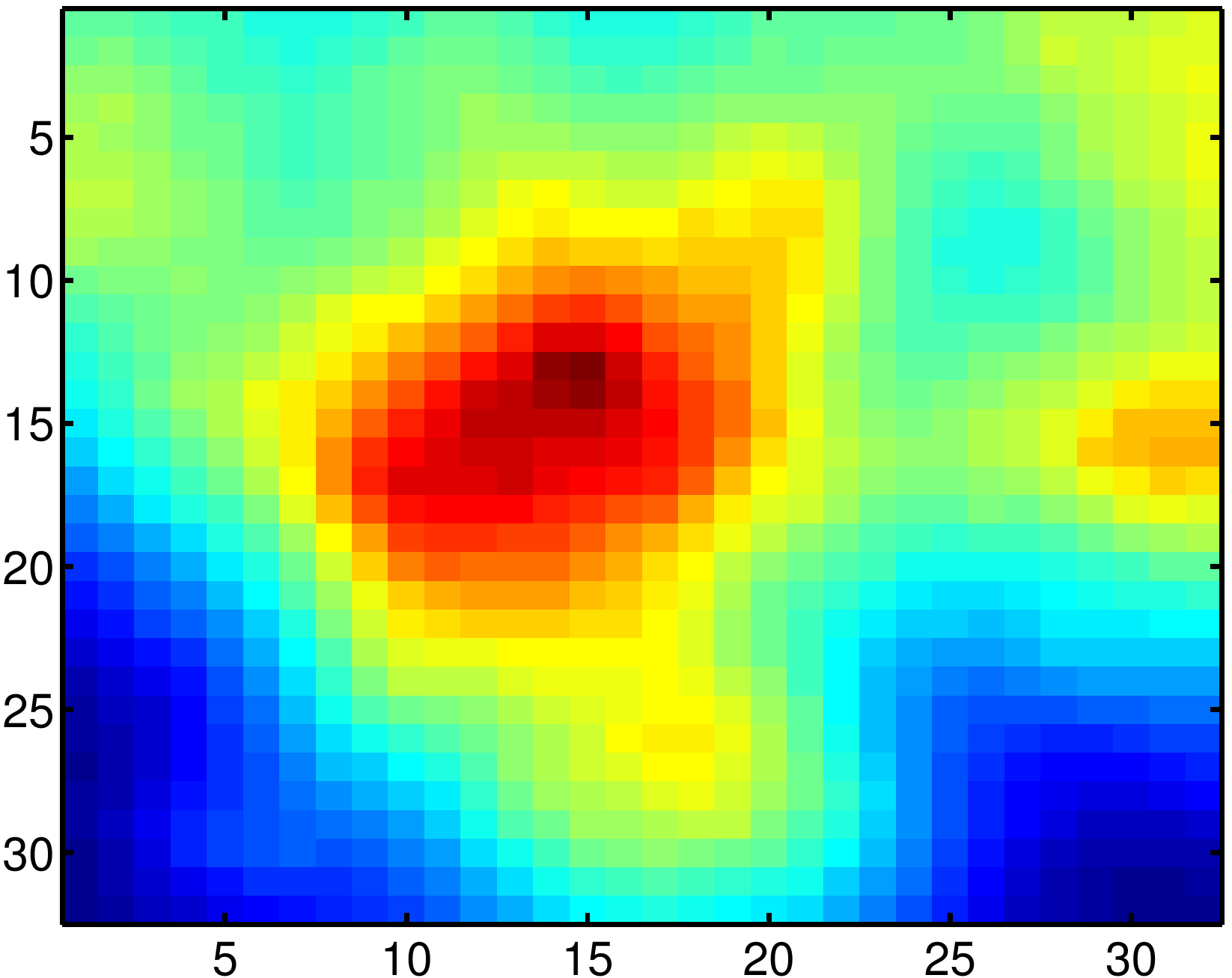}}
&
\resizebox{0.22\textwidth}{!}{\includegraphics{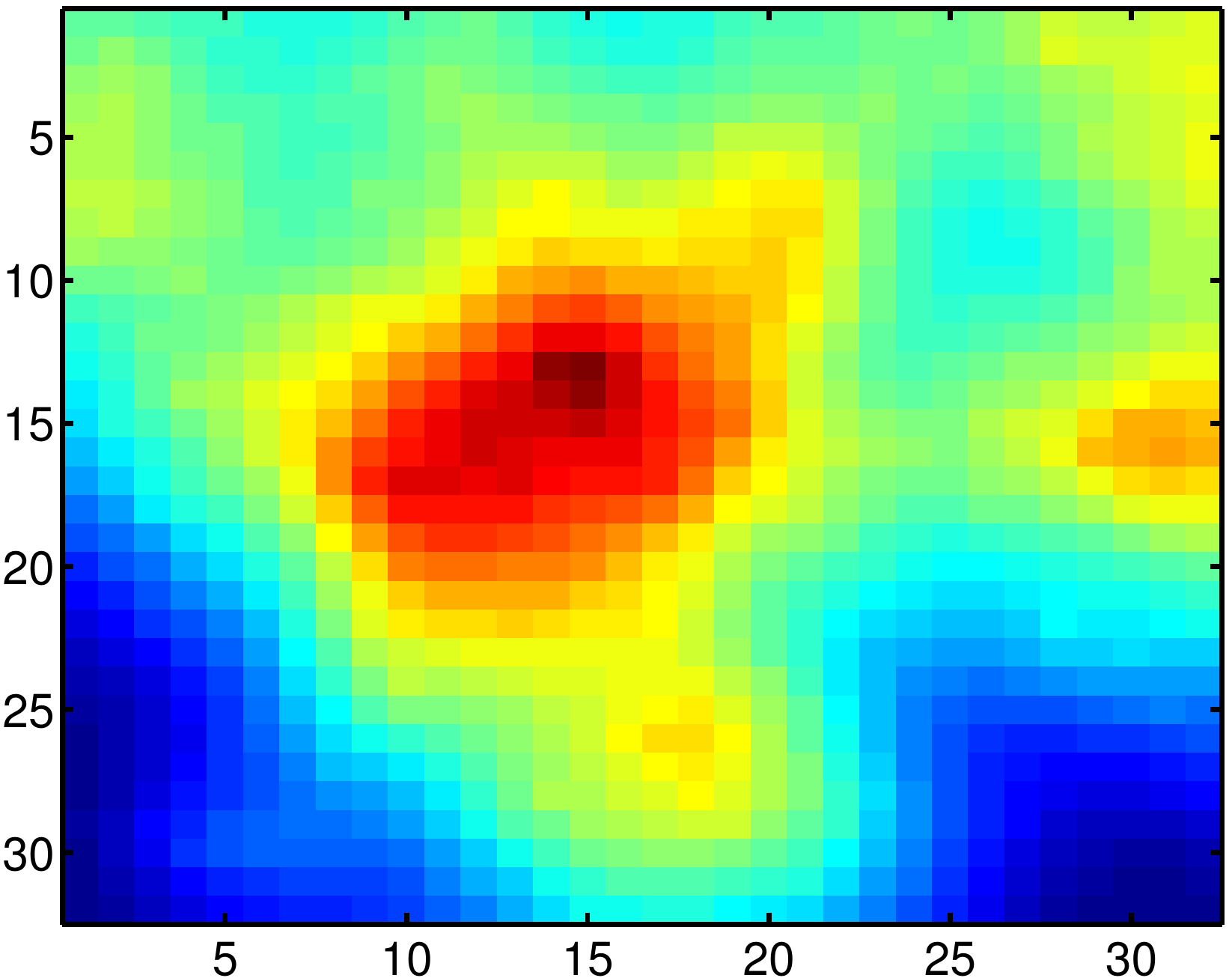}}
&
\resizebox{0.22\textwidth}{!}{\includegraphics{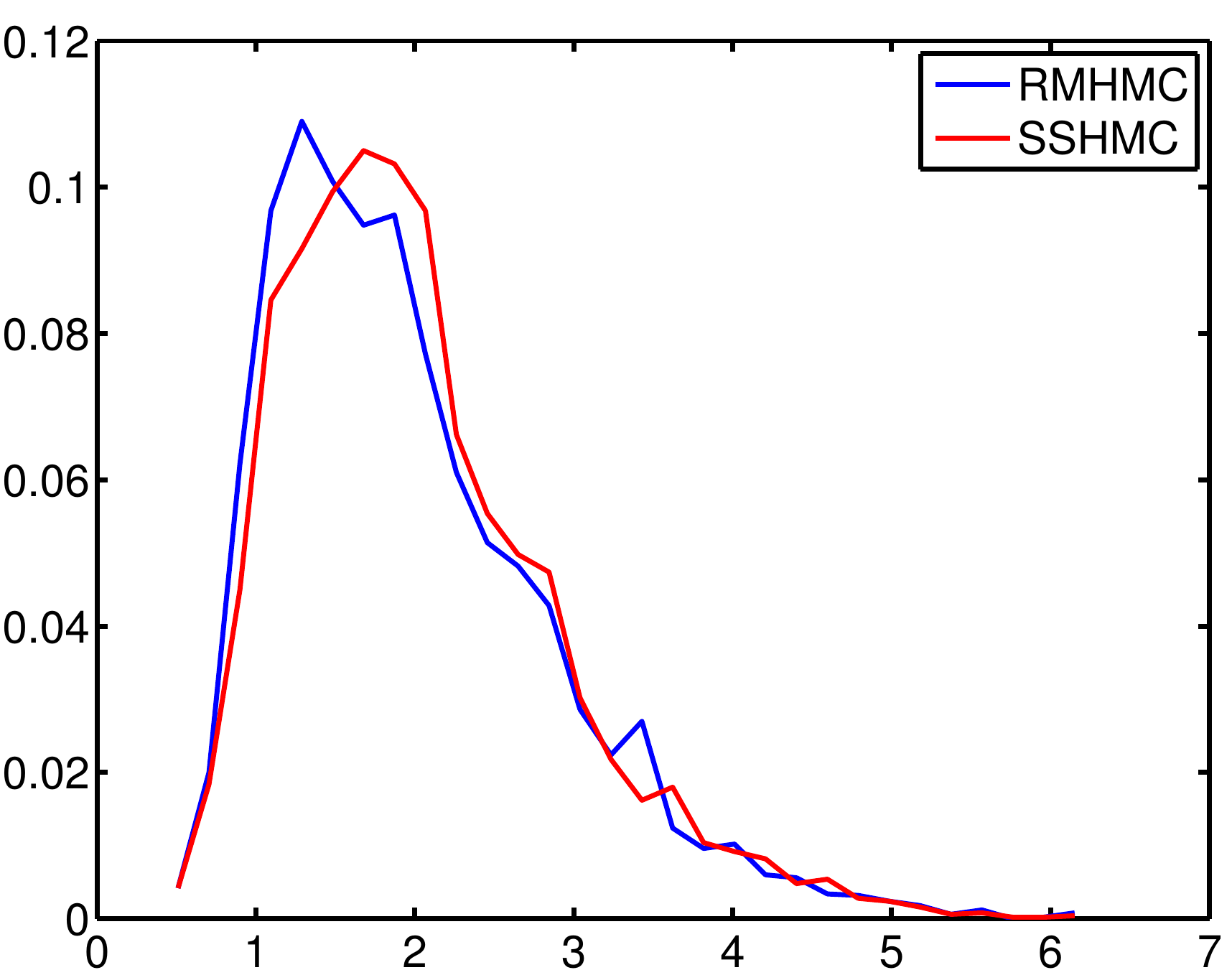}}
&
\resizebox{0.22\textwidth}{!}{\includegraphics{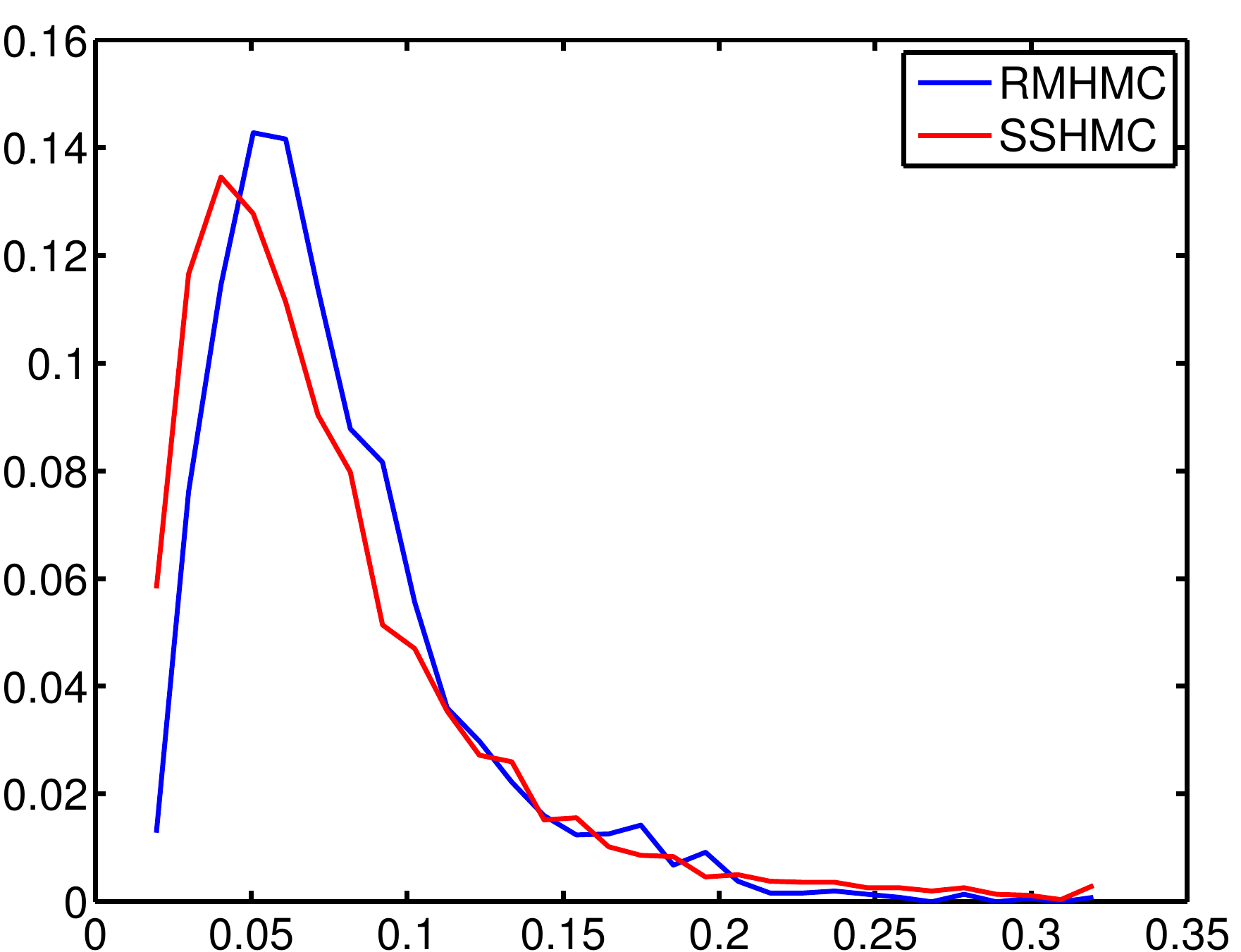}}

\end{tabular}
\caption{Sample mean of latent fields using RMHMC (left 1) and SSHMC (left 2).
The normalized histogram of sampled hyperparameter $\sigma$ (right 1) and $\beta$ 
(right 2). We draw 5000 samples from both methods after 1000 burn-in. The true
hyperparameter values are $(\sigma = 1.9, \beta = 0.03)$.
}\label{LGCX}
\vspace{-5pt}
\end{center}

\end{figure}

The results of ESS are given in Table \ref{tab:ESS_LGCox}.
The mean of sampled latent variable and the histogram of sampled hyperparameters 
are given in Figure \ref{LGCX}.
It is clear that the samples of RMHMC and SSHMC are consistent, so
both methods are mixing well.
However, SSHMC generates about \emph{six times} as many effective samples per hour as
RMHMC within Gibbs.

\begin{table}[tb]
	\centering
	\begin{tabular}{lllll}
		 & time(h) & ESS $\bx$(min, med, max) & ESS($\sigma, \beta$) &
		 min ESS/h\\
		 \hline
		 \sfmhmcf\ & 2.6&(\textbf{7.8}, \textbf{30},
		 \textbf{39})$\times10^2$ &(\textbf{2101},
		 \textbf{270})&\textbf{103.8}\\
		RMHMC(Gibbs) & 2.64&(1, 29, 38.3)$\times10^2$ & (200, 46)&
		16\\
		\hline
	\end{tabular}
	\caption{The ESS of 5000 posterior samples from 32x32 LGCPP after 1000
	burn-in samples. $\bx$ is the 1024 dimensional vector of latent variables and
	($\sigma, \beta$) are the hyperparameters of the Gaussian Process prior.
	``min ESS/h'' means minimum ESS per hour.}
		\label{tab:ESS_LGCox}
		\vspace{-15pt}
\end{table}


\section{Conclusion}

We have presented Semi-Separable Hamiltonian Monte Carlo (SSHMC),
a new version of Riemannian manifold Hamiltonian Monte 
Carlo (RMHMC) that aims to retain the flexibility 
of RMHMC for difficult Bayesian sampling problems,
while achieving greater simplicity and lower computational
complexity.  We tested SSHMC
on several different hierarchical models,
and on all the models we considered,
SSHMC outperforms both HMC and RMHMC within Gibbs 
in terms of number of effective samples
produced in a fixed amount of computation time.
Future work could consider
other choices of mass matrix within the semi-separable
framework, or the use of SSHMC within discrete models,
following previous work in discrete HMC \cite{zhang12,pakman13}.




\newpage
{\small
\bibliographystyle{abbrvnat}
\bibliography{database}
}

\end{document}